# Network nestedness in primates: a structural constraint or a biological advantage of social complexity?

Maxime Herbrich[1*], Eythan Cousin[1,2*], Ivan Puga-Gonzalez[3], Barbara Tiddi[4], Claudia Fichtel[5,6], Meg Crofoot[7,8,9,10], Andrew JJ MacIntosh[11,12], Erica van de Waal[13,14,15], Cédric Sueur[1,116]

1 Université de Strasbourg, IPHC UMR7178, CNRS, Strasbourg, France

2 Animal Behavior & Cognition, Department of Biology, Utrecht University, The Netherlands

3 Institute for Religion, Philosophy and History, University of Agder, Kristiansand, Norway

4 Agriculture, Health, and Environment Dept., Natural Resources Institute, University of Greenwich, UK

5 Behavioral Ecology and Sociobiology Unit, German Primate Center, Leibniz Institute for Primate Research, Göttingen, Germany.

6 Leibniz Science Campus "Primate Cognition", Göttingen 37077, Germany.

7 Department for the Ecology of Animal Societies, Max Planck Institute of Konstanz, Germany

8 Department of Biology, University of Konstanz, Konstanz, Germany

9 Center for the Advanced Study of Collective Behavior, University of Konstanz, Konstanz, Germany

10 Smithsonian Tropical Research Institute, Ancon, Panama

11 Primate Research Institute, Kyoto University, Inuyama, Japan

12 Wildlife Research Center, Kyoto University, Kanrin, Inuyama, Japan

13 Department of Ecology and Evolution, University of Lausanne, Lausanne, Switzerland

14 The Sense Innovation and Research Center, Lausanne and Sion, Switzerland

15 Inkawu Vervet Project, Mawana Game Reserve, Swart Mfolozi, KwaZulu Natal, South Africa

16 Institut Universitaire de France, Paris, France

* First coauthors

**Abstract:** This study investigates the prevalence and implications of nestedness within primate social networks, examining its relationship with cognitive and structural factors. We analysed data from 51 primate groups across 21 species, employing network analysis to evaluate nestedness and its correlation with modularity, neocortex ratio, and group size. We used Bayesian mixed effects modelling to investigate nestedness in primate social networks, controlling for phylogenetic dependencies and exploring various factors like neocortex ratio and group size. Our findings reveal a significant occurrence of nestedness in 66% of the species studied, exceeding chance expectations. This nestedness was more pronounced in groups with less steep dominance hierarchies, contrary to traditional assumptions linking it to hierarchical social structures. A notable inverse relationship between nestedness and modularity was observed, suggesting a structural trade-off in network formation. This pattern persisted even after controlling for species-specific social behaviours, indicating a general structural feature of primate networks. Surprisingly, our analysis showed no significant correlation between nestedness and neocortex ratio or group size, challenging the social brain hypothesis and suggesting a greater role for ecological factors in cognitive evolution. This study emphasises the importance of weak links in maintaining network resilience. Overall, our research provides new insights into primate social network structures, highlighting complex interplays between network characteristics and challenging existing paradigms in cognitive and evolutionary biology.

## 1. Introduction

Network science, foundational to modern complexity science, examines recurring structures in diverse systems, from biology to technology. These 'scale-free' or 'modular' patterns, observed in systems ranging from protein networks to ecosystems, offer evolutionary advantages like optimal communication (Oltvai & Barabási, 2002; Barabasi et al., 2003; Ravasz & Barabási, 2003; Clune et al., 2013; Marcoux & Lusseau, 2013; Ravasz et al., 2002; Romano et al., 2018; Sueur, 2023).

Nestedness is a key network characteristic, where fewer connected nodes (specialists) are embedded within the networks of more connected ones (generalists). This concept, studied primarily in bipartite ecological systems, indicates how specialists share interactions with generalists, akin to Russian nesting dolls. High nestedness is common in ecological networks, such as food webs and plant-pollinator interactions (Cantor et al., 2017). Cantor et al. (2017) extended the study of nestedness to unipartite biological networks, analysing species like Guiana dolphins, bottlenose dolphins, and spotted hyenas. They found unexpectedly high nestedness, influenced by social structures like fission-fusion dynamics in dolphins and dominance hierarchies in hyenas. This suggests a compromise between interaction specificity and affinity in network formation. Nestedness in animal social networks also indicates cognitive challenges, particularly in maintaining group cohesion in complex structures like fractal patterns with multiple modularity levels where subgroups are recursively nested within larger subgroups. Studies show a correlation between network efficiency and cognitive ability, with higher modularity in primate networks than theoretical models, hinting at an evolutionary advantage in disease mitigation and cognitive capability (Dunbar & Shultz, 2021; Amici et al., 2008; Pasquaretta et al., 2014; Romano et al., 2018, 2021). In summary, network science reveals crucial insights into the structures and functions of complex systems, highlighting the role of nestedness and modularity in biological networks and their connection to cognitive capacities and evolutionary advantages in animal societies.

The social brain hypothesis posits that social complexity drives cognitive evolution, where complexity refers to the differentiation of social relationships (Dunbar, 1998; Moscovice et al., 2020). Bergman & Beehner (2015) suggest quantifying this through the number of differentiated relationships based on rank, kinship, age, etc., while Fischer et al. (2017) propose measuring the diversity of relationships using four indices. In group-living animals, individuals navigate social dynamics for resource access and well-being, forming relationships that vary over time, often spanning their entire life (Barrett & Henzi, 2006; Schino & Aureli, 2017; Seyfarth, 1977). Social strategies within these groups can be complex. In hierarchically structured societies, such as despotic ones, social interactions and benefits are unevenly distributed (Matsumura, 1999; Sueur et al., 2011; Byrne & Whiten, 1989). Individuals adopt specialist or generalist roles based on factors like kinship or age, shaping their social networks. Specialists tend to have focused interactions and smaller networks, often fulfilling specific roles like caregivers. Generalists, conversely, engage more broadly and diversely, forming less specialised but more extensive networks (Sosa et al., 2021). Cantor et al. (2017) introduce the concept of network nestedness, which balances the quantity and quality of connections, as a measure of social relationship differentiation and an indicator of social complexity. This nestedness reflects the degree to which individuals or species vary from generalist to specialist roles. The social brain hypothesis further links larger brain sizes to greater flexibility in social interactions and relationship management. Larger brains are thought to enable more strategic behaviour and a wider range of social interactions (Dunbar, 1998). However, quantifying the number and diversity of relationships across species is challenging due to data limitations. Thus, nestedness serves as a comprehensive metric, encapsulating both the frequency and diversity of an individual's relationships within their social group.

Exploring primate social networks, we predict variations in nestedness influenced by cognitive abilities, group dynamics, and social structures. Key predictions include:

1. Negative Relationship Between Nestedness, Neocortex Ratio, and Group Size: A higher neocortex ratio, indicative of advanced cognitive abilities, likely leads to more strategic social interactions and the formation of specific social bonds. This increased cognitive capacity allows for selective social partnerships, which might reduce nestedness in social networks (Bissonnette et al., 2009; Chapais, 1995; Gilby et al., 2013; J. B. Silk et al., 2004; Barrett & Henzi, 2005; Dunbar, 1998; Dunbar & Shultz, 2021). Furthermore, larger group sizes offer diverse social connections, potentially decreasing nestedness as individuals form less intertwined relationships.
2. High Nestedness in Species with Despotic Societies: Species with lower neocortex ratios, often seen in despotic societies, are expected to show higher nestedness. Here, social interactions are concentrated among fewer individuals, resulting in more nested social networks.
3. Exception in Groups with High Male/Female Sex Ratios: An exception is predicted in groups with high male-to-female ratios. In such scenarios, the competition for limited mating opportunities may lead to less nested social networks, as males strategically select partners, forming specific alliances.
4. Interdependence of Network Properties: Our fourth prediction considers the complex interplay between network properties such as nestedness, modularity, and density. These properties are not entirely independent and can be influenced by evolutionary processes or structural constraints. Studies indicate that nestedness and modularity can co-occur under certain conditions, influenced by factors like disease prevention strategies in primate networks (Valverde et al., 2020; Fortuna et al., 2010; Romano et al., 2018, 2021). This interdependence underscores the complexity of network dynamics and the need to consider multiple factors when analysing nestedness in social networks.

This study aims to decipher the factors affecting nestedness in animal social networks, focusing on cognitive capabilities, group size, social organisation, and sex ratios. Utilising a dataset of socio-positive interactions from 38 primate groups and 21 species, we constructed and simulated social networks to calculate nestedness metrics. These simulations allowed controlled manipulation of network aspects, isolating variables to determine if observed patterns are due to external factors or inherent network structures. We investigated how network structure changes affect properties like nestedness and modularity. Our analysis incorporated network density, considering its impact on network metrics (Rankin et al., 2016), and examined group size and neocortex ratio, as larger brains are hypothesised to correlate with more complex social strategies (Dunbar & Shultz, 2021; Kudo & Dunbar, 2001; Lehmann et al., 2007). Additionally, we evaluated sex ratios to understand varying social strategies (Barrett & Henzi, 2006; Smuts & Smuts, 1993). By comparing empirical and simulated data while acknowledging the interplay between network properties (Fortuna et al., 2010; Valverde et al., 2020), we explored whether modularity and nestedness relationships are structural or involve evolutionary processes (Sueur, 2023; Sueur et al., 2019, Romano et al., 2018). This comprehensive approach sheds light on the complex dynamics of animal social networks.

## 2. Methods

### 2.1. Empirical data

We analysed 51 primate groups (Table 1, SI). The dataset comprised 21 species representing 5 families: 2 species of Atelidae, 3 species of Cebidae, 10 species of Cercopithecidae, 4 species of Hominidae, and 2 species of Lemuridae. Sex ratio was calculated including all individuals. The data was obtained from previously published studies (see Table S1, SI for references). Most of the data were freely available in accessible papers, while other data were utilized with the authors' permission. Group size ranged from 5 to 58 individuals. To generate social networks, we used matrices of socio-positive interactions (body contacts including social grooming, or proximities). Interaction type had no effect on the values of nestedness (Welch Two Sample t-test: body contact (n=38) vs proximity (n=13); values original matrix, t=0.02, p-value>0.05; values after cut-off 1, t=-2.31, p-value>0.05). In all cases, social networks were weighted and symmetrised; although they were later binarised to compute the values of nestedness (see below). The values of the neocortex ratio were taken from (Kudo & Dunbar, 2001; Stephan et al., 1981). We did not include phylopatric sex as variable as most of the studied species (n=49) are female-philopatric.

### 2.2. Social Network measures

#### 2.2.1. Nestedness

Nestedness is a network metric that calculates the degree to which the interactions of the least specialist individuals are nested within the interactions of the more generalist individuals (Cantor et al., 2017). A network is said to be nested when individuals show a hierarchical pattern in the way they interact (Fig 1). The value of nestedness indicates the degree to which specialist individuals share common interaction partners with the more generalist ones.

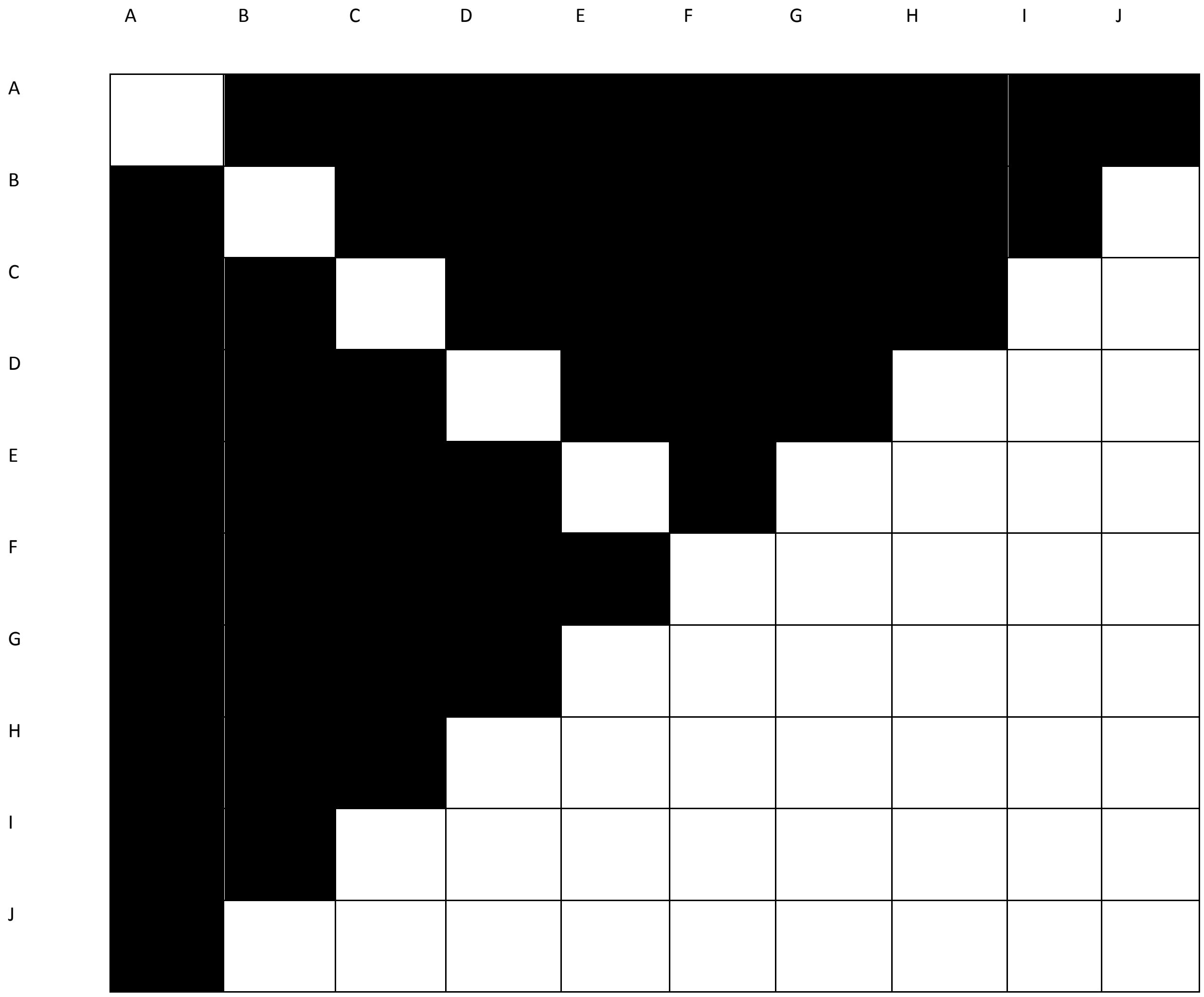

**Figure 1.** A perfectly nested matrix of interactions (n=10 individuals) would be represented by the following staircase-like pattern: from the most specialist individual J (bottom left) interacting with only one partner, to the most generalist one A (top right) interacting with all group members. Black filled = interaction, white filled = no interaction.

Nestedness is a typical measure used with bipartite networks (i.e. networks with two different kinds of nodes, e.g. plant-pollinator networks) and it is usually calculated using the Nestedness metric that is based on the Overlap and Decreasing Fill –NODF – method (Cantor et al., 2017). Recently, a modified version of the NODF method was developed to calculate nestedness in uni-partite networks: the UNODF method (Cantor et al. 2017). Like the NODF, the UNODF method is based on the properties of paired overlap and decreasing fill. The algorithm considers a binary symmetric matrix of interactions with n rows and n columns (n= # of individuals). The overlap in interactions with other group members by any given pair of individuals x and y is computed according to the relative position of the paired individuals in the rows of the matrix, and the degree (amount of interaction partners) of each paired individual: Kx and Ky. If the position of individual x in the matrix is at a higher row than that of individual y (rowx > rowy) and the degree of individual x is greater than that of individual y (Kx > Ky); then, the pair of individuals conforms to the notion of decreasing fill DFxy = 1. On the contrary; if Kx ≤ ky then DFxy = 0. The paired overlap between x and y (POxy) is then calculated as the proportion of interactions individual x shares with individual y. The degree of paired nestedness Nxy = 0 if DFxy = 0 and Nxy = POxy if DFxy = 1. UNODF is computed as the average of Nxy according to the following equation.

$$UNODF = \frac{2}{N(N-1)} \sum_{i}^{N} \sum_{j,i<j}^{N} \frac{\sum_{l \neq i,j}^{N} \left(1 - \delta_{k_i k_j}\right) A_{il} A_{jl}}{\min\left(k_i k_j\right)}$$

Where N is the total number of individuals in the network; A is the N × N binary symmetric matrix with all elements (ail) represented by rows (i) and columns (l); Ki and Kj are the degrees of individual i and j respectively; and δKiKj is the Kronecker delta with value 1 if ki = kj, and 0 otherwise. In this way, it discards the overlap between individuals with the same degree (ki = kj) and computes only the proportion of the interactions of the individual with the smaller degree that are also present in the set of interactions of the individual with larger degree (Cantor et al. 2017). In completely non-nested networks, UNODF = 0. For instance, fully connected networks (density = 1) have 0 nestedness since all individuals have the same degree. On the other hand, in perfectly nested networks (Fig 1), the UNODF will approach but never reach 1 because interactions of element i with itself (aii = 0) are disregarded (Cantor et al., 2017).

The calculation of nestedness was performed on binary symmetrical matrices. To account for the effect of weighted links in the networks of our data set (n=51), from the original adjacency matrix of each network, we created another 10 different binary symmetric matrices by filtering out links according to successive link weight thresholds (cut-off values) (Cantor et al., 2017). To do so, first we divided each original matrix by its maximum link weight; hence, we standardised the matrix by creating weight values ranging between 0 and 1 (i.e. ail ∈ [0,1]) (Fig 1, SI). Then, we defined 10 cut-offs at 0.1 intervals going from 0.0 to 0.9 and filter out links for each cut-off. Thus, for a cut-off of 0.1, links' weights below 0.1 were filtered out: if ail < 0.1, then ail= 0; if ail ≥ 0.1, then ail = 1 (Fig 1. SI). For each of these new matrices, the UNODF value was recalculated. In this way, according to the different cut-offs, we filtered out casual relationships (low weight values) from meaningful ones (high weight values). This way of filtering allowed us to check the importance of the weight of the links, from weak to strong (see Granovetter, 1973 for theory about 'the strength of weak ties').

#### 2.2.1.1 Significance tests for nestedness

By means of the UNODF method, we obtained a value of nestedness for each network and each of the network's cut-offs. These values were compared to the distribution of nestedness values obtained from theoretical random networks. In this way, we assessed if the value observed was significantly higher (lying on the right side) or lower (lying on the left side) than the distribution of values obtained from random networks. To create the random networks, we followed the same methodology as Cantor et al. (2017) where the probability of a link connecting two individuals is proportional to the number of links observed in both individuals. In the random networks the number of individuals, density of the network, and heterogeneity in degree distribution are held the same as in the original network. For each empirical network (the original and the resulting networks from each cut-off), 1000 random networks were built. Nestedness was calculated from the full-observation data set or after the removal of the weakest interactions to account for frequency of interactions. To assess the statistical significance of the observed value of nestedness, we compared the observed value with a null distribution of values of nestedness obtained by permuting the empirical network 1000 times. Nestedness was considered significantly higher (or lower) whenever the observed value lay on the right (left) side of the 95% confidence intervals of the random distribution.

All analyses were performed using R software v 3.3.2. The R code to calculate the nestedness metric, and that to create the null models, is available as supplementary material.

### 2.2.2. Density

The density of each network was computed as the number of observed edges divided by the number of possible edges (n2-n where n is equal to group size). Neither direction nor weights of interaction are taken into account.

### 2.2.3. Modularity

This metric represents the difference between the proportion of the total association of individuals within clusters (i.e. subgroups) and the expected proportion, given the summed associations of the different individuals. Modularity was calculated using igraph R package (Csardi & Nepusz, 2006); the igraph function is based on the eigenvector centrality of each individual and the method developed by (Newman, 2006). We also calculated the modularity based on betweenness centrality (Sosa et al., 2020; 2021). A high value of modularity means a high number of contacts within a subgroup but few contacts between subgroups, and low modularity means a homogeneous distribution of contacts between all group members. Furthermore, because modularity is positively correlated with centralisation (Pasquaretta et al., 2014), networks that are highly modular are also highly centralised and vice versa.

## 2.3. Phylogenetic and statistical analyses

### 2.3.1. The Primate Phylogeny

Given that one of our objectives was to identify the factors influencing the value of nestedness, it was necessary to model the relationship between the dependent variable (nestedness) and the various explanatory factors. However, since our data originated from different primate species within evolutionarily related families, phylogenetic relatedness could have an impact on the observed nestedness values. Therefore, we needed to control for phylogenetic dependency by incorporating the classification of our primates into the modelling process.

To achieve this, we utilised version 3 of the 10kTrees website (http://10ktrees.nunn-lab.org; (Arnold et al., 2010)) to construct a consensus tree summarising the topology and branch lengths of the trees obtained through the Bayesian MC3 method (Arnold et al., 2010). Additionally, to ensure consistency in species names between our data and the classification, we adopted the primate taxonomy presented in the GenBank DNA sequence database.

#### 2.3.2. Statistical Models

In our study, we utilised Bayesian mixed effects modelling via the BRMS package within R version 4.3.0 to analyse primate social networks, focusing on the nestedness within these networks across 51 populations of 21 species. This method incorporated phylogeny and multiple populations, allowing us to examine interspecies and intraspecies variations (Bürkner, 2017; R Core Team, 2013). We employed a zero-inflated Beta distribution for the response variable, considering nestedness values ranged between 0 and 1, with many groups showing a nestedness of 0 (Bürkner, 2019). Our calibration involved running four Markov chains with 125,000 iterations each, ensuring proper model convergence and appropriate mixing of the chains. This process was informed by literature recommendations (Gelman et al., 2013; J. Kruschke, 2014; J. K. Kruschke, 2010). We explored factors potentially associated with nestedness, such as neocortex ratio, group size, sex ratio, density, and modularity, using a comprehensive model that included all variables and their combinations (Cantor et al., 2017; Kudo & Dunbar, 2001; Lehmann et al., 2007; Moscovice et al., 2020; Pasquaretta et al., 2014; Smuts & Smuts, 1993). Model selection was based on the Leave-One-Out information criterion (LOOIC), akin to the Akaike information criterion (AIC), which assesses predictive accuracy for new data (Vehtari et al., 2017). We ranked models based on their Expected Log Predictive Density (ELPD) value and selected the best-performing models, considering the difference in ELPD and the associated standard error. In cases where the null model was a candidate, indicating no association between predictors and the response variable, the selection was based purely on statistical analysis. Our findings revealed several models with similar explanatory power for the data, forming a group distinctly different from others. To address this, we used the stacking method, integrating models based on their contribution to explaining nestedness (Hollenbach & Montgomery, 2020). The final model selection involved examining the posterior mean and credibility interval of the variables, determining the nature and significance of their relationships with nestedness.

### 2.4. Testing the meaning of the link between nestedness and modularity.

In our study, we explored the relationship between network nestedness and modularity, investigating whether this correlation is a structural constraint within primate social networks. Our analysis involved two primary methods:

- Comparison of Regression Slopes: We compared the regression slopes between modularity and nestedness in empirical (reduced from 51 to 38 due to null values in some networks) and simulated networks (N = 41,800). Simulated networks, designed to match the size and density of empirical networks, included two sets: one with 100 simulations per network and another with 1,000, to check for autocorrelation effects. By examining the similarity of slopes between empirical and simulated networks, we aimed to determine if the correlation between modularity and nestedness is a structural feature. If slopes are significantly different, it could indicate a preference or constraint for either modularity or nestedness.

- Assessment of Residual Variance: We evaluated the variance of residuals from the regression model nestedness~modularity, comparing empirical and simulated networks. A lower variance in residuals for empirical networks might indicate biological constraints, implying more consistent network properties compared to simulated ones. This analysis was conducted using linear regression and a Generative additive model (GAM), with linear regression generally showing better explanatory power (Marra & Wood, 2012; Wood, 2013). The variances were compared using an F test for both eigenvector and betweenness modularity.

These analyses aimed to discern whether the observed patterns in primate social networks are due to inherent structural constraints or biological factors influencing network properties. The findings contribute to our understanding of the factors shaping the complex dynamics within these social networks.

## 3. Results

### 3.1. Significance of Nestedness

Many primate groups and species (25 out of 51) had a value of nestedness significantly higher than the distribution of values obtained from the random networks (Fig 2; Table 1 in Supporting Information). Cut-offs had an influence on the detection of nestedness. Table 1 shows that networks from cut-off 0 (original network) or cut-off 1 (links with weight values below 0.1 filtered out), had a higher rate of occurrence of values of nestedness that were significantly higher than the random distribution: 29.41% for no cut-off and 27.45% for cut-off 1.

It should be noted that in all primate groups, after the cut-off value <0.6 no significant values of nestedness higher than the random distribution was detected (Table 2 SI). A probable explanation for this is that if too many links are filtered out (links with values <70% of the maximum); then it is not possible to deduce a hierarchical structure in the interactions of individuals. Further, four primate groups: one of S. apella, one of M. arctoides, one of M. radiatta and one of H. sapiens seem to be robust to different cut-off values as nestedness was detected up to cut-off value of 0.6 (Table 2 SI, data are proximities or body contacts).

Given that from networks of cut-offs 0 and 1 we obtained the highest rate of significant values of nestedness (Table2), our following analysis were performed only on networks from these two cut-offs.

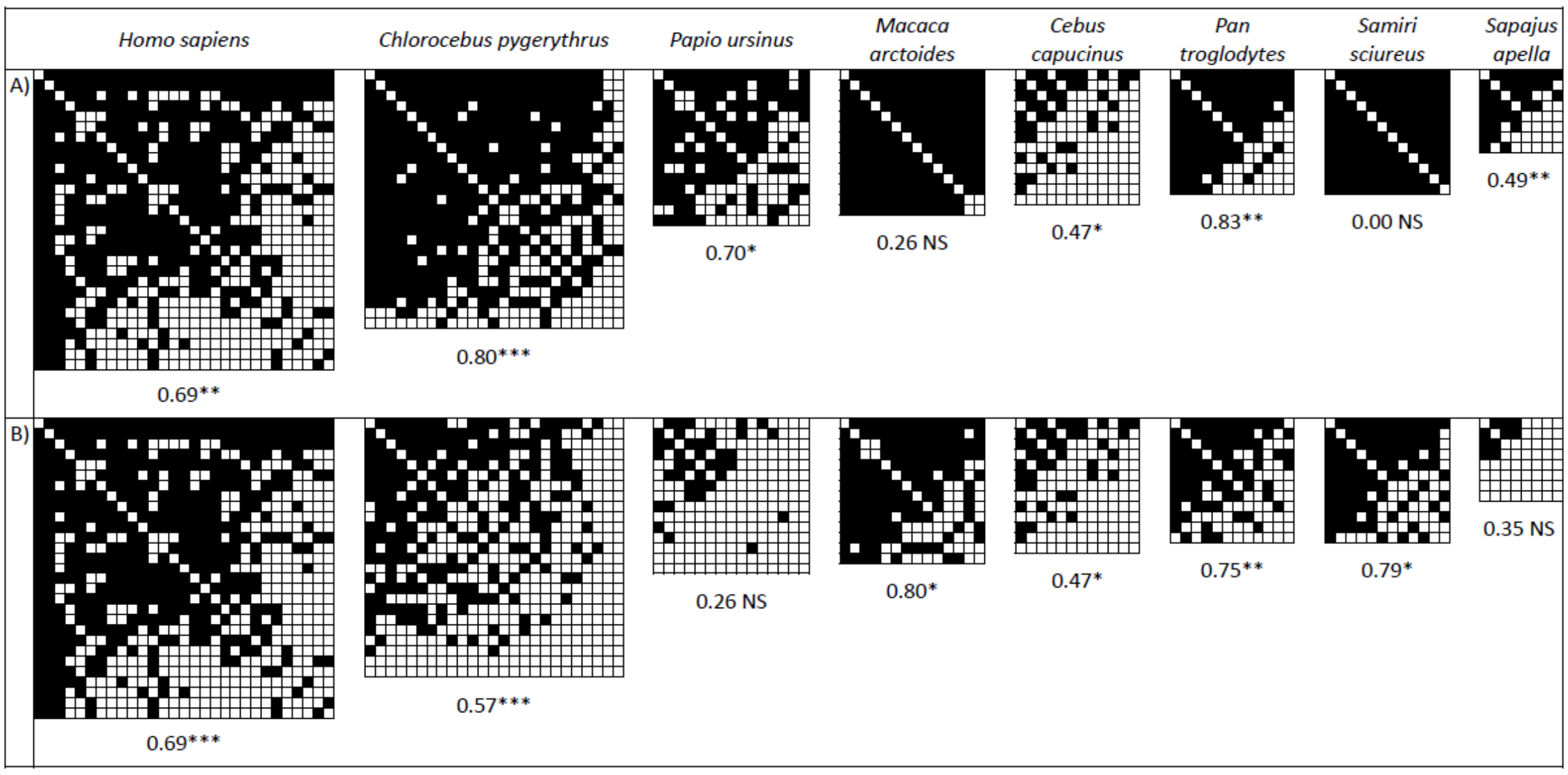

**Figure 2.** Matrices of different primate groups and their value of nestedness. Rows and columns represent different individuals, black squares a connection between two individuals and white squares no connection. Panel A) are original matrices, cut-off 0; and panel B) are matrices after cut-off 1 (link values <0.1 filtered out). Nestedness values are shown below each matrix. Significance p-values are based on 1000 permutations: NS= not significant; *=<0.05; **=<0.01; ***=<0.001.

| Cut-off value | % of values significantly higher | % of values significantly lower |
|---|---|---|
| 0 | 29.41 | 3.92 |
| <0.1 | 27.45 | 3.92 |
| <0.2 | 17.65 | 1.96 |
| <0.3 | 17.65 | 1.96 |
| <0.4 | 15.69 | 1.96 |
| <0.5 | 13.73 | 1.96 |
| <0.6 | 3.92 | 3.92 |
| <0.7 | 0.00 | 3.92 |
| <0.8 | 0.00 | 3.92 |
| <0.9 | 0.00 | 3.92 |

**Table 1.** Occurrence rate of values of nestedness significantly higher/lower than the random distribution according to each cut-offs. The p-values of each group and each cut-off are shown in tables 2 and 3 in Supporting information.

### 3.2. Bayesian models and selection

We studied the relationship between the nestedness value obtained from cut-off 1 (i.e. links weight values <0.1 were filtered out from the network) and the independent variables while controlling for the phylogenetic dependence of our data. Comparison of the different models using the loo information criterion (LOOIC) enabled us to create a statistical hierarchy of these models (Table 2).

| Variables included in the model | ELPD difference | Standard error of the difference |
|---|---|---|
| **Modularity + Density** | **0.0** | **0.0** |
| Modularity + Density + Neocortex | -0.3 | 0.8 |
| Modularity | -3.2 | 2.8 |
| Modularity + Neocortex | -3.6 | 2.8 |
| All | -5.1 | 2.2 |
| Null | -32.1 | 7.1 |

**Table 2.** Summary of the differences in ELPD and associated standard errors of the best models between the nestedness value obtained and the different independent variables. The differences are obtained from the best model (in bold). The grey line corresponds to the only model whose difference is not significant with the best model.

The model with the largest ELPD (smallest LOOIC) compared to our Bayesian model settings contains only modularity and density as significant explanatory variables. However, the model containing in addition the neocortex explanatory variable is not significantly different from the first model. As a result, even if modularity and density appear to be a better-fitting model than the previous one, it is not possible to conclude on the selection of one or the other from this approach unless we use the principle of parsimony. Nevertheless, this methodology enabled us to isolate two statistical models from a larger set of combinations.

We then considered a second method for statistically concluding our questioning. This includes our different models in a more global approach and distributes a weight according to the influence of each component on the general model.

| Variables included in the model | Weight |
|---|---|
| **Modularity + Density** | **0.971** |
| Modularity + Density + Neocortex | 0.029 |

**Table 3.** Weight distribution between the two selected models (total weight is 1). In bold, the best model according to the stacking method.

The stacking weights are largely concentrated on the modularity and density model, even though, according to the previous model selection (table 2), the two models are virtually equivalent. The model containing only the two variables therefore appears to have the best predictive accuracy and is the one we decided to retain.

Before looking in more detail at the variables in this model, we measured the phylogenetic signal (conventionally represented by the parameter $\lambda$) of our consensus phylogenetic tree estimated in this model. This was 0.14 and appeared to be significantly different from 0. However, as the credibility interval was extremely wide (0.00 - 0.94), it was difficult to really quantify the effect of phylogeny in the model, but we knew that it had an influence on our result and that phylogenetic modelling was necessary.

For our selected model, interspecies modularity, density, and intraspecies modularity had a significant effect on our nesting values (post. mean = – 2.94, CI = -5.25 to -0.76; post. mean = 1.86, CI = 0.68 to 3.15; post. mean = -5.15, CI = -7.08 to -3.36 respectively) (Fig.2). However, intraspecies density showed a non-significant positive relationship with our response variable (post. mean = 0.49, CI = -0.38 to 1.37).

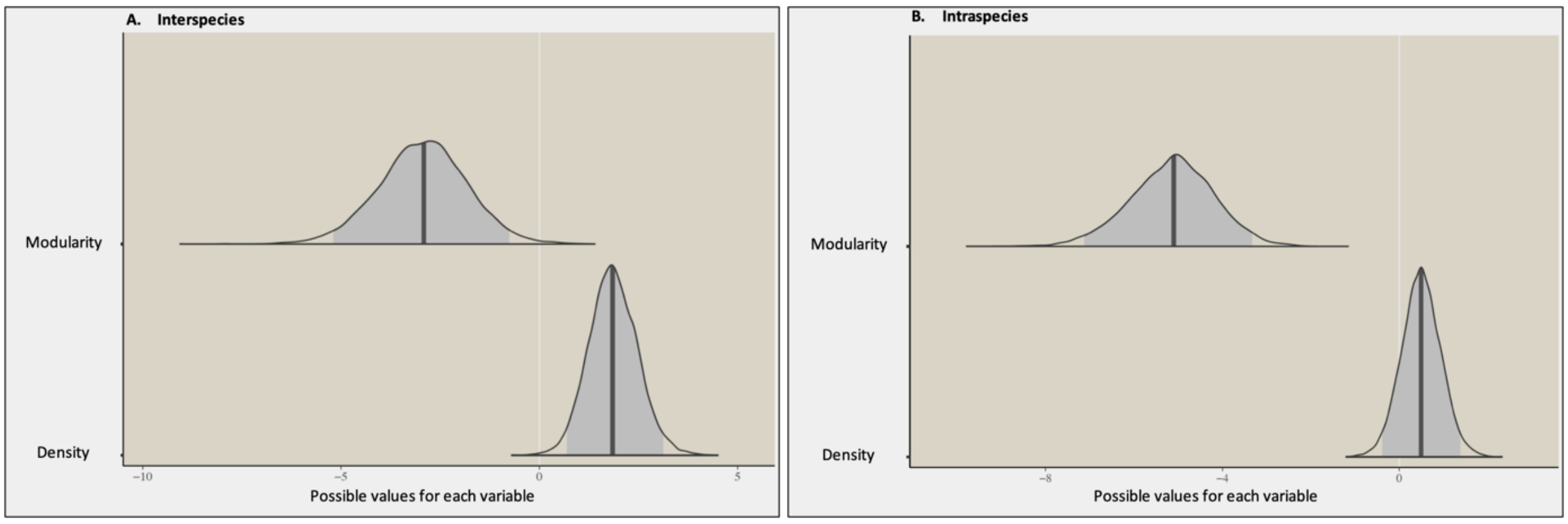

**Figure 3.** A posteriori distribution of the variables in the selected model (interspecies graph on the left and intraspecies graph on the right). The entire distribution curve corresponds to the probabilities of the different values that each variable can take (dark grey outline). The light grey area below the curve represents the 95% credibility interval and the vertical bar the posterior mean.

### 3.3. Testing the meaning of the link between nestedness and modularity.

We aim now to assess whether the link between modularity and nestedness is a structural constraint (inherent to network properties) or the result of an evolutionary process. In order to test this, we derived simulated networks from our empirical dataset. The 41800 simulated networks showed significant correlations between nestedness and modularity in 97.37% for betweenness modularity and in 100% for eigenvector modularity. This percentage decreases respectively to 81.58% and 18.42% when considering only correlations with an $r^2>0.1$. The simulated networks showed a $r^2=0.74$ ($p<0.0001$, fig.4a) for betweenness modularity and nestedness; whereas there was an $r^2=0.9$ ($p<0.0001$, fig.4b) for the 38 permuted empirical networks. Concerning eigenvector modularity and nestedness, we found a $r^2 = 0.75$ ($p<0.0001$, fig.4c) for the simulated networks; while the one of empirical networks was 0.87 ($p<0.0001$, fig.4d). The slopes of the relation between modularity and nestedness are not different between the empirical networks and the simulated networks, whatever the modularity metric (Student test; $t< 1.96$ for $\alpha=0.05$ and $ddl/\alpha=\infty$; fig. 5).

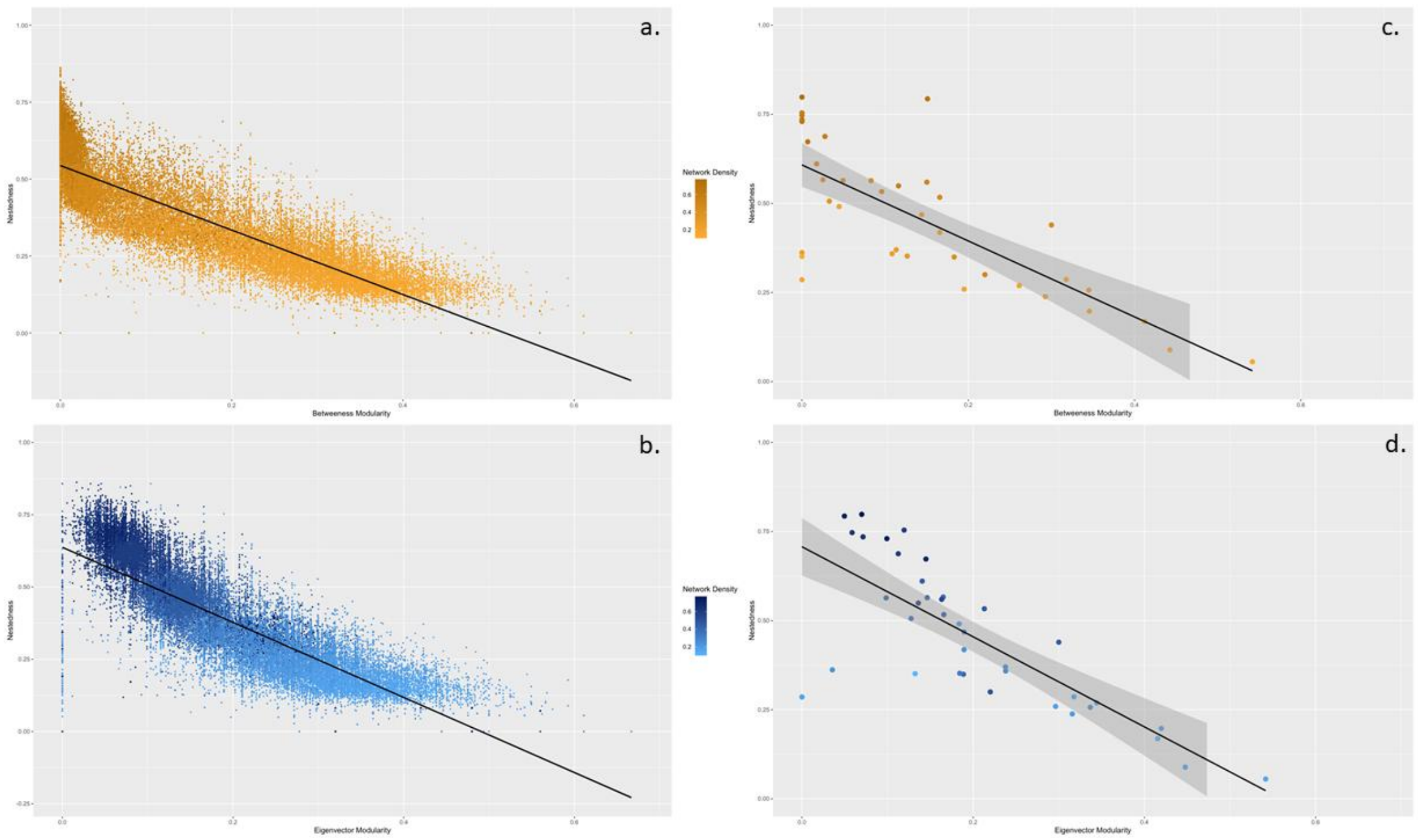

**Figure 4.** Correlation between nestedness and modularity (a) (top left) betweenness modularity for the simulated networks; Adjusted $r^2$ of the regression line: 0.73. (b) (bottom left) for eigenvector modularity for the simulated networks; Adjusted $r^2$ of the regression line: 0.70. (c) (top right) betweenness modularity for the 38 empirical networks. Adj $r^2$ = 0.90 (d) (bottom right) eigenvector modularity for the 38 empirical networks. Adj $r^2$ = 0.87

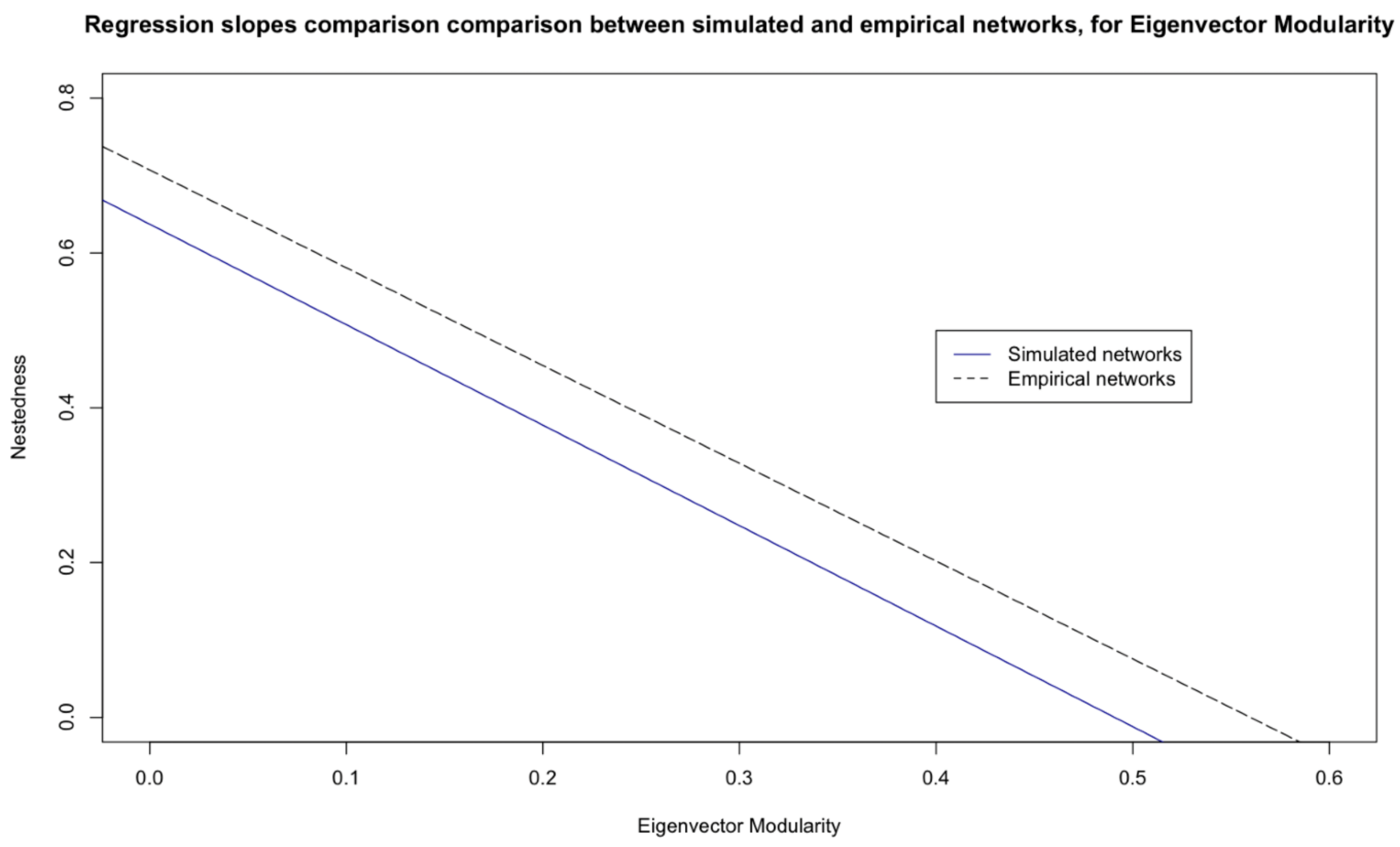

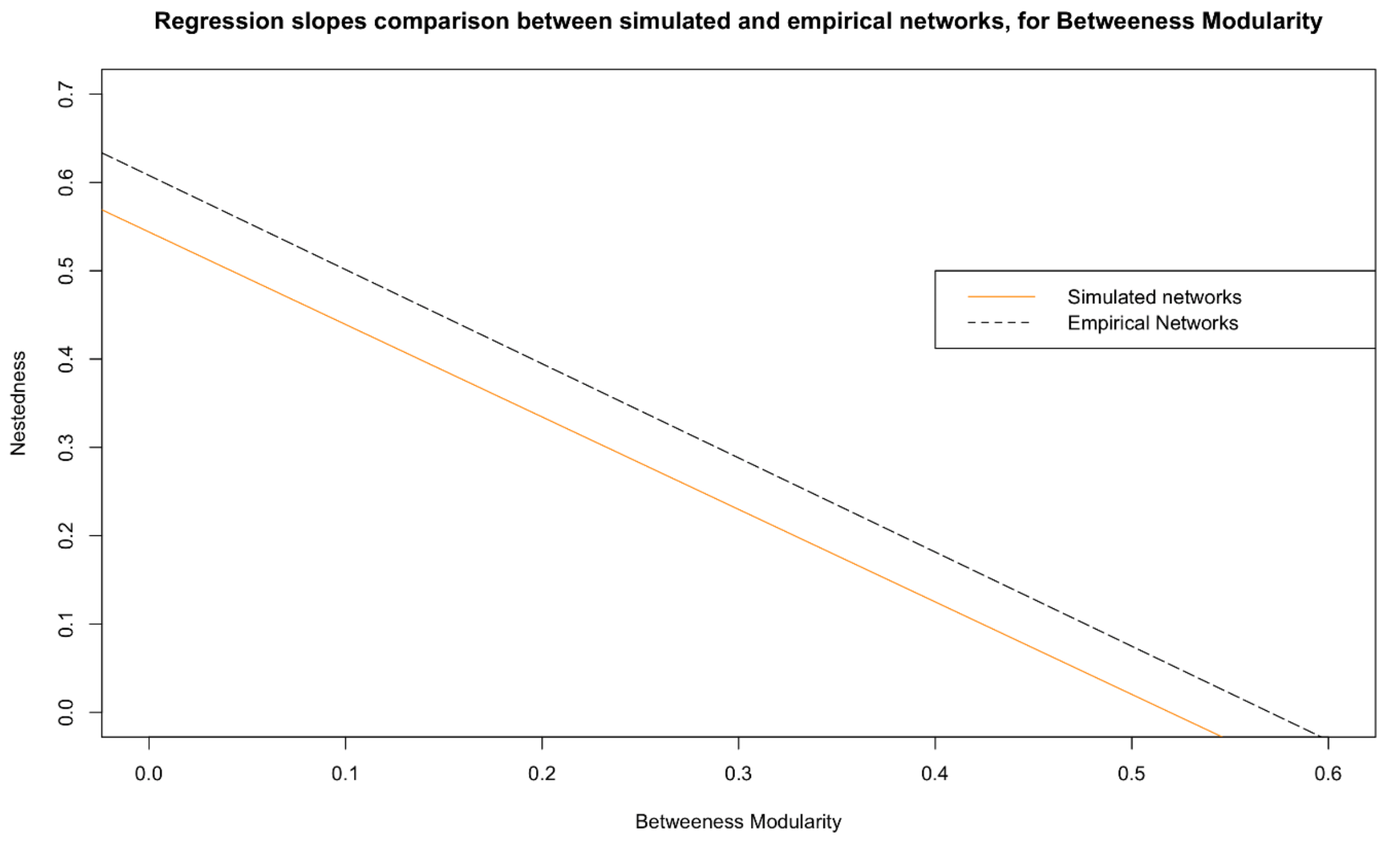

**Figure 5.** Comparison of the linear regression slopes between simulated networks (coloured) and empirical networks (black) for betweenness modularity (orange) and eigenvector modularity (blue).

Concerning the analyses of residuals, we found that empirical networks showed significant differences (F test for each simulated versus empirical network, $F<0.47$, $p <0.05$) to simulated networks concerning the betweenness modularity in 58.5% (51.9% with empirical data variance being higher than the simulated one, and 6.5% the inverse) and concerning the eigenvector modularity in 23.4% (19.9% with empirical data variance being higher than the simulated one, and 3.5% the inverse).

## 4. Discussion

In our study, we analysed social interaction distribution among group members in primates using the network metric of nestedness. We found that 49% of groups and 66% of species exhibited nestedness levels exceeding chance expectations. Typically, the social contacts of less social individuals (specialists) overlapped with those of more social ones (generalists). Additionally, phylogeny influenced social interaction patterns, suggesting that shared evolutionary traits might partly explain nestedness variability. However, quantifying this variation is challenging due to high uncertainty. Furthermore, network-specific parameters like modularity and density showed correlations, positive or negative, with nestedness, indicating interdependencies among these variables that may reflect differences in interspecies social styles and intraspecies personalities.

### 4.1 Biological implications of the relation between nestedness and modularity

In our study, we investigated the phenomenon of nestedness in uni-partite primate networks, a trait commonly associated with bipartite networks. Our results showed that 66% of the primate species studied exhibited higher-than-expected nestedness values. This widespread occurrence of nestedness in various biological organisations suggests potential advantages for primate groups, such as resilience against perturbations or disease prevention (Cantor et al., 2017; Sah et al., 2017; Puga-Gonzalez et al., 2019). Contrary to expectations, higher nestedness was observed in groups with less steep dominance hierarchies. Typically, in societies with pronounced hierarchies, individuals engage in grooming and commodity exchange with high-ranking members, leading to nested social interactions among similar ranks. However, our findings showed that increased modularity, which indicates relationship clustering, corresponded with decreased nestedness. This pattern persisted even when New World monkeys and strepsirrhines, known for commodities exchange competition, were excluded from the analysis (Port et al., 2009; Lazaro-Perea et al., 2004; Tiddi et al., 2011; Talbot et al., 2011).

To determine if the relationship between nestedness and modularity was an evolutionary trait or a structural network constraint, we compared empirical and simulated networks. The similarity in regression slopes and residual patterns between these networks suggested that the relationship is a structural constraint, not influenced by individual behaviours. This conclusion aligns with findings in host-pathogen, pollination, and host-parasite plant networks, where a negative correlation between modularity and nestedness indicates a trade-off between these metrics (Romano et al., 2018, 2020, 2021). Our study highlights the complexity of primate social structures, suggesting that while modularity may facilitate information transmission and disease control, nestedness could reflect hierarchical structures with implications for group dynamics. This trade-off between modularity and nestedness underscores the structural limitations within animal social networks.

### 4.2 Nestedness and neocortex ratio

Contrary to expectations derived from the social brain hypothesis, our study found no correlation between nestedness and neocortex ratio in primate societies. The hypothesis posits that complex social living drives the development of cognitive abilities for strategic relationship formation beyond just dominance rank, incorporating factors like age, kinship, and third-party relationships (Dunbar, 1998; Bergman & Beehner, 2015; Fischer et al., 2017). We hypothesised a negative correlation between neocortex ratio and nestedness, expecting more strategic alliance formation, but our results did not support this. Several factors might explain this lack of correlation. Firstly, the benefits of cooperating within kin groups could outweigh those with non-kin, leading individuals to maintain kin-centric relationships. Secondly, the biological market theory suggests that hierarchy steepness influences competition and commodities exchange in primate societies, affecting relationship choices (Barrett & Henzi, 2006; Noë & Hammerstein, 1995). In shallow hierarchies, alliances may not significantly impact payoffs, whereas in steep hierarchies, kinship alliances could be more beneficial than non-kinship ones. Thirdly, certain network properties may offer advantages like rapid information access, with network modularity and centralisation impacting information transfer efficiency (Pasquaretta et al., 2014). Networks in hunter-gatherer societies, for instance, are structured for efficient transmission of cumulative culture (Migliano et al., 2017). However, the specific advantages conferred by nestedness or how network structures may constrain individual social strategies remain largely unexplored. Additionally, it could be hypothesized that nestedness is more commonly found in societies with shallower hierarchies, where the need for support and coalition, thus creating modularity, is generally smaller. In such societies, bigger social tolerance might occur alongside this dynamic. Consequently, social relationships could be less constrained by the need for support, potentially allowing individual personality traits to play a more significant role in the development of relationships. This dynamic might lead to more curious or bolder individuals developing a broader range of social relationships, becoming 'generalists' in social interactions, as opposed to 'shyer' individuals who might become more specialized, leading to the creation of nested structures. This suggests a fascinating potential interplay between individual personality traits and social structure in primate societies.

### 4.3. Nestedness and group size

Our study found no correlation between nestedness and neocortex ratio, nor with group size (a proxy for social complexity), challenging previous findings (Lehmann & Dunbar, 2009; Barton, 1996; R. I. Dunbar, 1992; Pérez-Barbería et al., 2007; Shultz & Dunbar, 2007). A linear regression analysis, controlling for phylogenetic signal, revealed no significant relationship between group size and neocortex ratio (PLGS: neocortex ratio ~ log group size: n=21, F= 0.00, p-value = 0.984). This contradicts several past studies but aligns with recent research suggesting ecological factors, such as diet and home range size, are better predictors of neocortex size than group size (DeCasien et al., 2017; Powell et al., 2017). These findings indicate ecological factors may be more influential than social complexity in brain size evolution.

Regarding sex ratio, we hypothesised that male-biased groups would show strategic alliance formation, affecting nestedness. However, we found no correlation, possibly due to infrequent coalition formation or coalition choices based on dominance rather than strategic selection (Bissonnette et al., 2014; Hemelrijk & Puga-Gonzalez, 2012). Factors like reproductive skew and demographic composition might influence coalition behaviours, overriding assumed strategic choices (Bissonnette et al., 2009; Perry et al., 2004; Righi & Takacs, 2014). Our results contribute to the growing body of evidence questioning the link between neocortex ratio, group size, and social complexity, and highlight the complexity of coalition formation in primate societies.

### 4.4. Methodological aspects

Our study on primate social networks, limited by a small sample size, could not statistically test the impact of fission-fusion dynamics on nestedness. We analysed data from two species known for these dynamics: spider monkeys (Ateles geoffroyi) and chimpanzees (Pan troglodytes), with one spider monkey group and three chimpanzee groups studied in the wild. Their nestedness values, both in wild and captive groups, did not exceed the species' maximum or minimum values, with chimpanzees showing ~0.70 and 0.36, and spider monkeys at 0.36.

Consistent with Cantor et al. (2017), significant nestedness was often detected after removing weaker links from the network, highlighting the distinction between casual and meaningful relationships. While this method is crucial for certain analyses, weak links should not be dismissed as they can contribute to network robustness and resilience (Fushing et al., 2013; Pasquaretta et al., 2014; Puga-Gonzalez et al., 2019).

An ANOVA indicated species effects on nestedness values from original matrices, but not after link removal. Tukey post-hoc tests showed no significant differences, though some comparisons were borderline. This variation in nestedness among species could be due to differing group densities. Regression analysis revealed no correlation between nestedness and density, suggesting nestedness is independent of the number of network connections, varying based on how links are distributed among individuals.

## 5. Conclusion

This study reveals that nestedness, indicative of the extent to which fewer connected individuals share connections with more connected ones, is a prevalent feature in primate social networks. The observed interplay between nestedness and network modularity suggests a structural balance, highlighting a potential trade-off within these networks. While the benefits of modularity have been extensively studied, the specific advantages of nestedness in social networks remain underexplored. Future research should consider both these aspects in tandem to fully understand their combined impact on social dynamics and the evolutionary advantages they may offer to primate groups. This integrated approach could significantly enhance our comprehension of social network structures in primates.

## Data availability

Scripts and data are available at: https://doi.org/10.5281/zenodo.10674792

## Conflicts of Interest

The author declared no conflicts of interest

## Acknowledgments

We thank B. Thierry, N. Claidière, M. Shimada, A.J. King, P.M. Kappeler, C. Bret, P. Izar and S. Perry for sharing their data (from Pasquaretta et al. 2014 Sci Rep). We thank Lydia Ikhlef for her help on statistics.

## Funding

We acknowledge the support of S.P. Lambeth and S.J. Schapiro in collecting these chimpanzee data. The chimpanzee colony at UT MD Anderson Cancer Center is supported by NIH U42 (OD-011197). C.S. is granted by the University of Strasbourg Institute for Advanced Study (USIAS) and the Fyssen Foundation. E.W. and the vervet monkeys data were funded Sinergia grant (CRSI33_133040) from the Swiss National Science Foundation to R. Bshary, C. P. van Schaik and A.W. A.J.J.M. was supported by the Japan Ministry of Environment, Culture, Sports, Science and Technology (MEXT), the Japan Society for the Promotion of Sciences (JSPS) and the Cooperative Research Program of the Wildlife Research Center, Kyoto University.

**References**

Amici, F., Aureli, F., & Call, J. (2008). Fission-fusion dynamics, behavioral flexibility, and inhibitory control in primates. Current Biology, 18(18), 1415‑ 1419.

Apicella, C. L., Marlowe, F. W., Fowler, J. H., & Christakis, N. A. (2012). Social networks and cooperation in hunter-gatherers. Nature, 481(7382), 497‑ 501. https://doi.org/10.1038/nature10736

Arnold, C., Matthews, L. J., & Nunn, C. L. (2010). The 10kTrees website : A new online resource for primate phylogeny. Evolutionary Anthropology: Issues, News, and Reviews, 19(3), 114‑ 118.

Barabasi, A.-L., Dezso, Z., Ravasz, E., Yook, S.-H., & Oltvai, Z. (2003). Scale-Free and Hierarchical Structures in Complex Networks. In P. L. Garrido & J. Marro (Éds.), MODELING OF COMPLEX SYSTEMS: Seventh Granada Lectures (Vol. 661, p. 1‑ 16). AIP. https://doi.org/10.1063/1.1571285

Barrett, L., & Henzi, P. (2005). The social nature of primate cognition. Proceedings of the Royal Society B: Biological Sciences, 272(1575), 1865‑ 1875.

Barrett, L., & Henzi, S. P. (2006). Monkeys, markets and minds : Biological markets and primate sociality. In Cooperation in primates and humans (p. 209‑ 232). Springer.

Barton, R. A. (1996). Neocortex size and behavioural ecology in primates. Proceedings of the Royal Society of London. Series B: Biological Sciences, 263(1367), 173‑ 177.

Bergman, T. J., & Beehner, J. C. (2015). Measuring social complexity. Animal Behaviour, 103, 203‑ 209. https://doi.org/10.1016/j.anbehav.2015.02.018

Bissonnette, A., de Vries, H., & van Schaik, C. P. (2009). Coalitions in male Barbary macaques, Macaca sylvanus : Strength, success and rules of thumb. Animal Behaviour, 78(2), 329‑ 335.

Bissonnette, A., Franz, M., Schülke, O., & Ostner, J. (2014). Socioecology, but not cognition, predicts male coalitions across primates. Behavioral Ecology, 25(4), 794‑ 801. https://doi.org/10.1093/beheco/aru054

Bürkner, P.-C. (2017). brms : An R package for Bayesian multilevel models using Stan. Journal of statistical software, 80, 1‑ 28.

Bürkner, P.-C. (2019). Bayesian item response modelling in R with brms and Stan. arXiv preprint arXiv:1905.09501.

Byrne, R., & Whiten, A. (1989). Machiavellian Intelligence : Social Expertise and the Evolution of Intellect in Monkeys, Apes, and Humans (Oxford Science Publications). {Oxford University Press, USA}. http://www.amazon.ca/exec/obidos/redirect?tag=citeulike09-20&path=ASIN/0198521758

Cantor, M., Pires, M. M., Marquitti, F. M., Raimundo, R. L., Sebastián-González, E., Coltri, P. P., Perez, S. I., Barneche, D. R., Brandt, D. Y., & Nunes, K. (2017). Nestedness across biological scales. PloS one, 12(2), e0171691.

Chapais, B. (1995). Alliances as a means of competition in primates : Evolutionary, developmental, and cognitive aspects. American Journal of Physical Anthropology, 38(S21), 115‑ 136.

Clune, J., Mouret, J.-B., & Lipson, H. (2013). The evolutionary origins of modularity. Proceedings of the Royal Society B: Biological Sciences, 280(1755), 20122863. https://doi.org/10.1098/rspb.2012.2863

Csardi, G., & Nepusz, T. (2006). The igraph software package for complex network research. InterJournal, Complex Systems, 1695(5), 1‑ 9.

DeCasien, A. R., Williams, S. A., & Higham, J. P. (2017). Primate brain size is predicted by diet but not sociality. Nature Ecology & Evolution, 1(5), 0112.

Dunbar, R. (1998). The social brain hypothesis. Evolutionary Anthropology: Issues, News, and Reviews, 6(5), 178‑ 190.

Dunbar, R. I. (1992). Neocortex size as a constraint on group size in primates. Journal of human evolution, 22(6), 469‑ 493.
Dunbar, R. I., & Shultz, S. (2021). Social complexity and the fractal structure of group size in primate social evolution. Biological Reviews.
Fischer, J., Farnworth, M. S., Sennhenn-Reulen, H., & Hammerschmidt, K. (2017). Quantifying social complexity. Animal Behaviour, 130, 57‑ 66. https://doi.org/10.1016/j.anbehav.2017.06.003
Fortuna, M. A., Stouffer, D. B., Olesen, J. M., Jordano, P., Mouillot, D., Krasnov, B. R., Poulin, R., & Bascompte, J. (2010). Nestedness versus modularity in ecological networks : Two sides of the same coin? The Journal of Animal Ecology, 79(4), 811‑ 817. https://doi.org/10.1111/j.1365-2656.2010.01688.x
Franks, D. W., Ruxton, G. D., & James, R. (2010). Sampling animal association networks with the gambit of the group. Behavioral Ecology and Sociobiology, 64(3), 493‑ 503. https://doi.org/10.1007/s00265-009-0865-8
Fushing, H., Wang, H., VanderWaal, K., McCowan, B., & Koehl, P. (2013). Multi-Scale Clustering by Building a Robust and Self Correcting Ultrametric Topology on Data Points. PLoS ONE, 8(2), e56259. https://doi.org/10.1371/journal.pone.0056259
Gelman, A., Carlin, J. B., Stern, H. S., Dunson, D. B., Vehtari, A., & Rubin, D. B. (2013). Bayesian Data Analysis, Third Edition. CRC Press.
Gilby, I. C., Brent, L. J. N., Wroblewski, E. E., Rudicell, R. S., Hahn, B. H., Goodall, J., & Pusey, A. E. (2013). Fitness benefits of coalitionary aggression in male chimpanzees. Behavioral Ecology and Sociobiology, 67(3), 373‑ 381. https://doi.org/10.1007/s00265-012-1457-6
Granovetter, M. S. (1973). The Strength of Weak Ties. American Journal of Sociology, 78(6), 1360‑ 1380. https://doi.org/10.1086/225469
Hemelrijk, C. K., & Puga-Gonzalez, I. (2012). An individual-oriented model on the emergence of support in fights, its reciprocation and exchange. PloS one, 7(5), e37271.
Henzi, S., Barrett, L., Gaynor, D., Greeff, J., Weingrill, T., & Hill, R. (2003). Effect of resource competition on the long-term allocation of grooming by female baboons : Evaluating Seyfarth's model. Animal Behaviour, 66(5), 931‑ 938.
Hollenbach, F. M., & Montgomery, J. M. (2020). Bayesian model selection, model comparison, and model averaging. In The sage handbook of research methods in political science and international relations (p. 937‑ 960). Sage.
Isbell, L. A., & Young, T. P. (2002). Ecological models of female social relationships in primates : Similarities, disparities, and some directions for future clarity. Behaviour, 139(2), 177‑ 202. https://doi.org/10.1163/156853902760102645
Kruschke, J. (2014). Doing Bayesian data analysis : A tutorial with R, JAGS, and Stan.
Kruschke, J. K. (2010). Bayesian data analysis. Wiley Interdisciplinary Reviews: Cognitive Science, 1(5), 658‑ 676.
Kudo, H., & Dunbar, R. I. M. (2001). Neocortex size and social network size in primates. Animal Behaviour, 62(4), 711‑ 722. https://doi.org/10.1006/anbe.2001.1808
Lazaro-Perea, C., de Fátima Arruda, M., & Snowdon, C. T. (2004). Grooming as a reward ? Social function of grooming between females in cooperatively breeding marmosets. Animal Behaviour, 67(4), 627‑ 636.
Lehmann, J., & Dunbar, R. I. M. (2009). Network cohesion, group size and neocortex size in female-bonded Old World primates. Proceedings of the Royal Society B: Biological Sciences, 276(1677), 4417‑ 4422. https://doi.org/10.1098/rspb.2009.1409
Lehmann, J., Korstjens, A. H., & Dunbar, R. I. M. (2007). Group size, grooming and social cohesion in primates. Animal Behaviour, 74(6), 1617‑ 1629. https://doi.org/16/j.anbehav.2006.10.025
Macdonald, S., Schülke, O., & Ostner, J. (2013). The absence of grooming for rank-related benefits in female Assamese macaques (Macaca assamensis). International Journal of Primatology, 34(3), 571‑ 584.
Marcoux, M., & Lusseau, D. (2013). Network modularity promotes cooperation. Journal of Theoretical Biology, 324, 103‑ 108. https://doi.org/10.1016/j.jtbi.2012.12.012

Marra, G., & Wood, S. N. (2012). Coverage Properties of Confidence Intervals for Generalized Additive Model Components. Scandinavian Journal of Statistics, 39(1), 53‑74. https://doi.org/10.1111/j.1467-9469.2011.00760.x
Matheson, M. D., & Bernstein, I. S. (2000). Grooming, social bonding, and agonistic aiding in rhesus monkeys. American Journal of Primatology: Official Journal of the American Society of Primatologists, 51(3), 177‑186.
Matsumura, S. (1999). The Evolution of 'Egalitarian' and 'Despotic' Social Systems Among Macaques. Primates, 40(1), 23‑31.
Migliano, A. B., Page, A. E., Gómez-Gardeñes, J., Salali, G. D., Viguier, S., Dyble, M., Thompson, J., Chaudhary, N., Smith, D., Strods, J., Mace, R., Thomas, M. G., Latora, V., & Vinicius, L. (2017). Characterization of hunter-gatherer networks and implications for cumulative culture. Nature Human Behaviour, 1, 0043. https://doi.org/10.1038/s41562-016-0043
Moscovice, L. R., Sueur, C., & Aureli, F. (2020). How socio-ecological factors influence the differentiation of social relationships : An integrated conceptual framework. Biology Letters, 16(9), 20200384. https://doi.org/10.1098/rsbl.2020.0384
Newman, M. E. J. (2006). Modularity and community structure in networks. Proceedings of the National Academy of Sciences, 103(23), 8577‑8582. https://doi.org/10.1073/pnas.0601602103
Noë, R., & Hammerstein, P. (1995). Biological markets. Trends in Ecology & Evolution, 10(8), 336‑339.
Oltvai, Z. N., & Barabási, A.-L. (2002). Life's Complexity Pyramid. Science, 298(5594), 763‑764. https://doi.org/10.1126/science.1078563
Pasquaretta, C., Levé, M., Claidière, N., van de Waal, E., Whiten, A., MacIntosh, A. J. J., Pelé, M., Bergstrom, M. L., Borgeaud, C., Brosnan, S. F., Crofoot, M. C., Fedigan, L. M., Fichtel, C., Hopper, L. M., Mareno, M. C., Petit, O., Schnoell, A. V., di Sorrentino, E. P., Thierry, B., … Sueur, C. (2014). Social networks in primates : Smart and tolerant species have more efficient networks. Scientific Reports, 4(1), 7600. https://doi.org/10.1038/srep07600
Pérez-Barbería, F. J., Shultz, S., & Dunbar, R. I. (2007). Evidence for coevolution of sociality and relative brain size in three orders of mammals. Evolution, 61(12), 2811‑2821.
Perry, S., Barrett, H. C., & Manson, J. H. (2004). White-faced capuchin monkeys show triadic awareness in their choice of allies. Animal Behaviour, 67(1), 165‑170. https://doi.org/10.1016/j.anbehav.2003.04.005
Port, M., Clough, D., & Kappeler, P. M. (2009). Market effects offset the reciprocation of grooming in free-ranging redfronted lemurs, Eulemur fulvus rufus. Animal Behaviour, 77(1), 29‑36.
Powell, L. E., Isler, K., & Barton, R. A. (2017). Re-evaluating the link between brain size and behavioural ecology in primates. Proceedings of the Royal Society B: Biological Sciences, 284(1865), 20171765.
Puga-Gonzalez, I., Sosa, S., & Sueur, C. (2019). Social style and resilience of macaques' networks, a theoretical investigation. Primates, 60(3), 233‑246.
R Core Team, R. (2013). R: A language and environment for statistical computing.
Ravasz, E., & Barabási, A.-L. (2003). Hierarchical organisation in complex networks. Physical review E, 67(2), 026112.
Ravasz, E., Somera, A. L., Mongru, D. A., Oltvai, Z. N., & Barabási, A.-L. (2002). Hierarchical organisation of modularity in metabolic networks. science, 297(5586), 1551‑1555.
Righi, S., & Takacs, K. (2014). Triadic balance and closure as drivers of the evolution of cooperation. Social Simulation Conference.
Romano, V., MacIntosh, A. J. J., & Sueur, C. (2020). Stemming the Flow : Information, Infection, and Social Evolution. Trends in Ecology & Evolution, 35(10), 849‑853. https://doi.org/10.1016/j.tree.2020.07.004
Romano, V., Shen, M., Pansanel, J., MacIntosh, A. J. J., & Sueur, C. (2018). Social transmission in networks : Global efficiency peaks with intermediate levels of modularity. Behavioral Ecology and Sociobiology, 72(9), 154. https://doi.org/10.1007/s00265-018-2564-9

Romano, V., Sueur, C., & MacIntosh, A. J. J. (2021). The trade-off between information and pathogen transmission in animal societies. Oikos. https://doi.org/10.1111/oik.08290

Sah, P., Leu, S. T., Cross, P. C., Hudson, P. J., & Bansal, S. (2017). Unraveling the disease consequences and mechanisms of modular structure in animal social networks. Proceedings of the National Academy of Sciences, 201613616.

Schino, G., & Aureli, F. (2017). Reciprocity in group-living animals : Partner control versus partner choice. Biological Reviews, 92(2), 665‑ 672.

Seyfarth, R. M. (1977). A model of social grooming among adult female monkeys. Journal of theoretical Biology, 65(4), 671‑ 698. https://doi.org/10.1016/0022-5193(77)90015-7

Shultz, S., & Dunbar, R. I. (2007). The evolution of the social brain : Anthropoid primates contrast with other vertebrates. Proceedings of the Royal Society B: Biological Sciences, 274(1624), 2429‑ 2436.

Silk, J., Altmann, J., & Alberts, S. (2006). Social relationships among adult female baboons (papio cynocephalus) I. Variation in the strength of social bonds. Behavioral Ecology and Sociobiology, 61(2), 183‑ 195. https://doi.org/10.1007/s00265-006-0249-2

Silk, J. B., Alberts, S. C., & Altmann, J. (2004). Patterns of coalition formation by adult female baboons in Amboseli, Kenya. Animal Behaviour, 67(3), 573‑ 582. https://doi.org/10.1016/j.anbehav.2003.07.001

Smuts, B. B., & Smuts, R. W. (1993). Male aggression and sexual coercion of females in nonhuman primates and other mammals : Evidence and theoretical implications. Advances in the Study of Behavior, 22(22), 1‑ 63.

Sosa, S., Sueur, C., & Puga-Gonzalez, I. (2021). Network measures in animal social network analysis : Their strengths, limits, interpretations and uses. Methods in Ecology and Evolution, 12(1), 10‑ 21. https://doi.org/10.1111/2041-210X.13366

Stephan, H., Frahm, H., & Baron, G. (1981). New and revised data on volumes of brain structures in insectivores and primates. Folia Primatologica; International Journal of Primatology, 35(1), 1‑ 29.

Sterck, E. H. M., Watts, D. P., & van Schaik, C. P. (1997). The evolution of female social relationships in nonhuman primates. Behavioral Ecology and Sociobiology, 41(5), 291‑ 309. https://doi.org/10.1007/s002650050390

Sueur, C. (2023). Socioconnectomics : Connectomics Should Be Extended to Societies to Better Understand Evolutionary Processes. Sci, 5(1), Article 1. https://doi.org/10.3390/sci5010005

Sueur, C., Petit, O., De Marco, A., Jacobs, A. T., Watanabe, K., & Thierry, B. (2011). A comparative network analysis of social style in macaques. Animal Behaviour, 82(4), 845‑ 852. https://doi.org/10.1016/j.anbehav.2011.07.020

Sueur, C., Romano, V., Sosa, S., & Puga-Gonzalez, I. (2019). Mechanisms of network evolution : A focus on socioecological factors, intermediary mechanisms, and selection pressures. Primates, 1‑ 15.

Talbot, C. F., Freeman, H. D., Williams, L. E., & Brosnan, S. F. (2011). Squirrel monkeys' response to inequitable outcomes indicates a behavioural convergence within the primates. Biology letters, 7(5), 680‑ 682.

Thierry, B., Singh, M., & Kaumanns, W. (2004). Macaque societies : A model for the study of social organisation. Cambridge University Press.

Tiddi, B., Aureli, F., Polizzi di Sorrentino, E., Janson, C. H., & Schino, G. (2011). Grooming for tolerance ? Two mechanisms of exchange in wild tufted capuchin monkeys. Behavioral Ecology, 22(3), 663‑ 669.

Valverde, S., Vidiella, B., Montañez, R., Fraile, A., Sacristán, S., & García-Arenal, F. (2020). Coexistence of nestedness and modularity in host-pathogen infection networks. Nature Ecology & Evolution, 4(4), 568‑ 577. https://doi.org/10.1038/s41559-020-1130-9

Vehtari, A., Gelman, A., & Gabry, J. (2017). Practical Bayesian model evaluation using leave-one-out cross-validation and WAIC. Statistics and computing, 27, 1413‑ 1432.

Wood, S. N. (2013). On p-values for smooth components of an extended generalized additive model. Biometrika, 100(1), 221- 228. https://doi.org/10.1093/biomet/ass048